\documentclass[a4paper]{spie} 
\usepackage[]{graphicx}
\usepackage{amssymb,amsmath, psfrag}
\newcommand{\rnum}{\bf {R}}

\newcommand{\dd}[1]{\mathrm{d}\,#1}
\newcommand{\LT}{\mathcal{L}}

\newcommand{\Alt}{{\mathrm{Alt}}}

\newcommand{\VField}[1]{{\bf #1}} 
\newcommand{\ltl}{L^2_{\rm loc}}
\newcommand{\hcl}{H_{\rm loc}({\rm curl})}
\newcommand{\hdl}{H_{\rm loc}({\rm div})}
\newcommand{\hpl}{H^{1}_{\rm loc}}
\newcommand{\hclkz}{H_{{\rm loc}, k_z}({\rm curl})}
\newcommand{\hdlkz}{H_{{\rm loc}, kz}({\rm div})}
\newcommand{\hplkz}{H^{1}_{{\rm loc}, k_z}}

\newcommand{\hckzd}[1]{H_{#1, k_z}({\rm curl})}

\newcommand{\hpkzd}[1]{H^1_{#1, {k_z}}}

\def\squareforqed{\hbox{\rlap{$\sqcap$}$\sqcup$}}
\def\qed{\ifmmode\else\unskip\quad\fi\squareforqed}

\newtheorem{problem}{{\bfseries Problem}}
\newtheorem{remark}{{\bfseries Remark}}

\title{JCMmode: An Adaptive Finite Element Solver for the Computation of Leaky Modes} 
\author{Lin Zschiedrich\supit{a, b}, Sven Burger\supit{a, b}, Roland Klose\supit{a},
Achim Sch\"adle\supit{a},  and Frank Schmidt\supit{a, b}
\skiplinehalf
\supit{a} Zuse Institute Berlin (ZIB), Takustra{\ss}e 7, D-14195 Berlin, Germany \\
\supit{b} JCMwave GmbH, Haarer Stra{\ss}e 14a, D-85640 Putzbrunn, Germany
}

\authorinfo{Further author information: (Send correspondence to Lin Zschiedrich) \\
E-mail: zschiedrich@zib.de \\
URL: http://www.zib.de/nano-optics/ 
}

\begin{document} 
\maketitle 

%%%%%%%%%%%%%%%%%%%%%%%%%%%%%%%%%%%%%%%%%%%%%%%%%%%%%%%%%%%%% 
\begin{abstract}
We present our simulation tool JCMmode for calculating propagating modes of an optical waveguide. As ansatz functions we use higher order, vectorial elements (Nedelec elements, edge elements). Further we construct transparent boundary conditions to deal with leaky modes even for problems with inhomogeneous exterior domains as for integrated hollow core Arrow waveguides. We have implemented an error estimator which steers the adaptive mesh refinement. This allows the precise computation of singularities near the metal's corner of a Plasmon-Polariton waveguide even for irregular shaped metal films on a standard personal computer. 
\end{abstract}
%%%%%%%%%%%%%%%%%%%%%%%%%%%%%%%%%%%%%%%%%%%%%%%%%%%%%%%%%%%%%
\section{INTRODUCTION}
\label{sect:intro} 
The computation of propagating modes of an optical waveguide is one of the central tasks in the optical component design. In mathematical modeling this corresponds to a quadratic eigenvalue problem in the sought propagation constant $k_z$~\cite{Jin:93a}. Beyond ``true'' eigenmodes with finite energy in the cross section there exist so-called ``leaky modes'' which are solutions to Maxwell's equations but with typically increasing field intensity for a growing distance to the waveguide core~\cite{Petracek:2002a,Uranus:2004a}. These leaky modes must satisfy a further asymptotic boundary condition for large distances to the waveguide core. Analog to scattering problems, one demands that there is no energy transport from infinity within the cross section~\cite{SchmidtF:01b,Hohage:03a}. To bring this into a mathematical form, we split the cross section $\rnum^{2}$ into a bounded interior domain $\Omega_{\rm int}$ and an exterior domain $\Omega_{\rm ext}$, that is $\rnum^{2} = \Omega_{\rm int} \cup \Omega_{\rm ext}$. For a homogeneous exterior domain $\Omega_{\rm ext}$ (with constant permittivity and permeability) the correct asymptotic boundary condition is the well known Silver-M{\"u}ller condition~\cite{Monk:2003a}. Uranus and Hoekstra use a BGT-like transparent boundary condition based on this asymptotic boundary condition~\cite{Uranus:2004a}. Besides a poor convergence with the size of the computational domain, this asymptotic boundary condition is wrong for inhomogeneous exterior domains~\cite{SchmidtF:01b}. But, many waveguide structures are composed of layers with an immense lateral expansion compared to the waveguide core diameter. These structures are best modeled in the way that the layers reach infinity. To deal with such inhomogeneous exterior domains in a rigorous manner, Schmidt has proposed the pole condition concept for the definition of asymptotic boundary conditions~\cite{SchmidtF:01b,Hohage:03a}. We briefly introduce this concept in Section~\ref{sect:trueandleaky}. Further we show the connection of this concept to a modified PML method proposed by the authors~\cite{Hohage:03c}. In the Section~\ref{sect:finiteelements} we explain how to discretize the modified PML method and how to couple the transparent boundary condition with the interior finite element discretization. In the last section we demonstrate the ability of our method for challenging problems in modern optical waveguide design.

Alternatively to the modified PML method Schmidt has presented a numerical approach which is directly based on the pole condition. The authors will compare these two methods in a succeeding paper.   
\section{LIGHT PROPAGATION IN A WAVEGUIDE}
\label{sect:lightinwaveguide}
Starting from Maxwell's equations in a medium without sources and free currents and assuming time-harmonic dependence with angular frequency $\omega>0$ the electric and magnetic fields
\[
\VField{E}(x, y, z, t)  = \widetilde{\VField{E}}(x, y, z)e^{-i\omega \cdot t}, \;
\VField{H}(x, y, z, t) = \widetilde{\VField{H}}(x, y, z)e^{-i\omega \cdot t}, \;
\]
must satisfy 
\begin{eqnarray*}
\nabla \times \widetilde{\VField{E}} & = & i\omega \mu \widetilde{\VField{H}}, \quad
\nabla \cdot \epsilon \widetilde{\VField{E}} = 0, \\
\nabla \times \widetilde{\VField{H}} & = & -i\omega \epsilon \widetilde{\VField{E}}, \quad 
\nabla \cdot \mu \widetilde{\VField{H}} = 0. \\
\end{eqnarray*}
Here $\epsilon$ denotes the permittivity tensor and $\mu$ denotes the permeability tensor of the materials. In the following we drop the wiggles, so that $\widetilde{\VField{E}} \rightarrow \VField{E}$, $\widetilde{\VField{H}} \rightarrow \VField{H}$. From the equations above we then may derive (by direct substitution) the second order equation for the electric field
\begin{eqnarray*}
\nabla \times \mu^{-1} \nabla \times \VField{E} - \omega^2 \epsilon \VField{E} & = & 0, \\
\nabla \cdot \epsilon \VField{E} & = & 0.
\end{eqnarray*}  
A similar equation holds true for the magnetic field - one only must replace $\VField{E}$ by $\VField{H}$ and interchange $\epsilon$ and $\mu$. Observe that any solution to the first equation also meets the divergence condition (second equation). This is because $\nabla \cdot \nabla \times  = 0.$\\
To recover the underlying structure we rewrite these equations in differential form,
\begin{subequations}
\label{thefieldmaxequationsdf}
\begin{eqnarray}
d_{1} \mu^{-1}  d_{1} \VField{e} - \omega^2 \epsilon \VField{e} & = & 0, \\
d_{2}  \epsilon \VField{e} & = & 0.
\end{eqnarray}  
\end{subequations}
A reader not familiar with this calcalus may replace the exterior derivatives $d_{0}$, $d_{1}$, $d_{1}$ with classical differential operators, $d_{0} \rightarrow \nabla$, $d_{1} \rightarrow \nabla \times $ and $d_{2} \rightarrow \nabla \cdot$. Here, the electric field appears as a differential 1-form, $e = e_{x}dx+e_{y}dy+e_{z}dz,$ whereas the material tensors act -- from a more mathematical point of view -- as operators
\begin{eqnarray*}
\epsilon,\, \mu \; : \; \Alt^{1} \rightarrow \Alt^{2}.
\end{eqnarray*}   
In order to derive a weak formulation we define the following function spaces on the domain $\Omega = \rnum^3$ 
\begin{eqnarray*}
\hpl & = & \left \{\phi \in \Alt^{0} \, | \, \nabla \phi \in (\ltl)^3 \right \} \\
\hcl & = & \left \{\VField{e} \in \Alt^{1} \, | \, (e_{x}, e_{y}, e_{z}) \in (\ltl)^3, \; \nabla \times (e_{x}, e_{y}, e_{z})^{\mathrm T}  \in (\ltl)^3 \right \} \\
\hdl & = & \left \{\VField{d} \in \Alt^{2} \, | \, (d_{x}, d_{y}, d_{z}) \in (\ltl)^3, \; \nabla \cdot (d_{x}, d_{y}, d_{z})^{\mathrm T}  \in \ltl \right \} 
\end{eqnarray*}
The weak form to Equations~\eqref{thefieldmaxequationsdf} now reads
\begin{subequations}
\label{thefieldmaxequationsdfweak}
\begin{eqnarray}
\int_{\rnum^{3}} \left(  \mu^{-1} d_{1} \VField{e} \wedge d_{1} \overline{\VField{v}}  - 
\omega^{2} (\epsilon \VField{e}) \wedge \overline{\VField{v}} \right ) & = & 0 \\
\int_{\rnum^{3}}  (\epsilon \VField{e}) \wedge d_{0} \overline{p} & = & 0
\end{eqnarray}
\end{subequations}
for all $\VField{v} \in \hcl$ and $p \in \hpl$ with compact support. 

An optical waveguide is an invariant structure in one spatial direction which we assume to be the $z$~-~direction of a cartesian coordinate system. A propagating mode is a solution to the above time-harmonic Maxwell's equations
such that the electric field $\VField{E}$ depends harmonically on the spatial coordinate~$z$, 
\[
\VField{E}(x, y, z)  =  \widehat{\VField{E}}(x, y)e^{ik_z \cdot z}.
\]
Hence a propagating mode travels along the $z$-direction. The scalar quantity $k_z$ is called {\em propagation constant}. Let us denote by $\hclkz$ the subspace of fields in $\hcl$ which depends on $z$ as $\VField{e}(x, y, z) = \widehat{\VField{e}}(x, y, z) \exp(ik_{z}z)$. The spaces $\hplkz$ and $\hdlkz$ are defined accordingly. It is sufficient to restrict the variational problem~\eqref{thefieldmaxequationsdfweak} on the cross section~$z=0$. As mentioned in the introduction a propagating mode should not only solve Maxwell's equations but should also transport no energy within the cross section from infinity, that is it should be {\em purely outgoing} in the cross section. The precise definition of what {\em purely outgoing} means is given in the next section. The weak waveguide problem is summarized in the following Problem~\ref{weakwaveguideproblem}.
\begin{problem}[Weak Waveguide Problem]
\begin{figure}
  \begin{center}
    \begin{tabular}{c}
      \includegraphics[width=10cm]{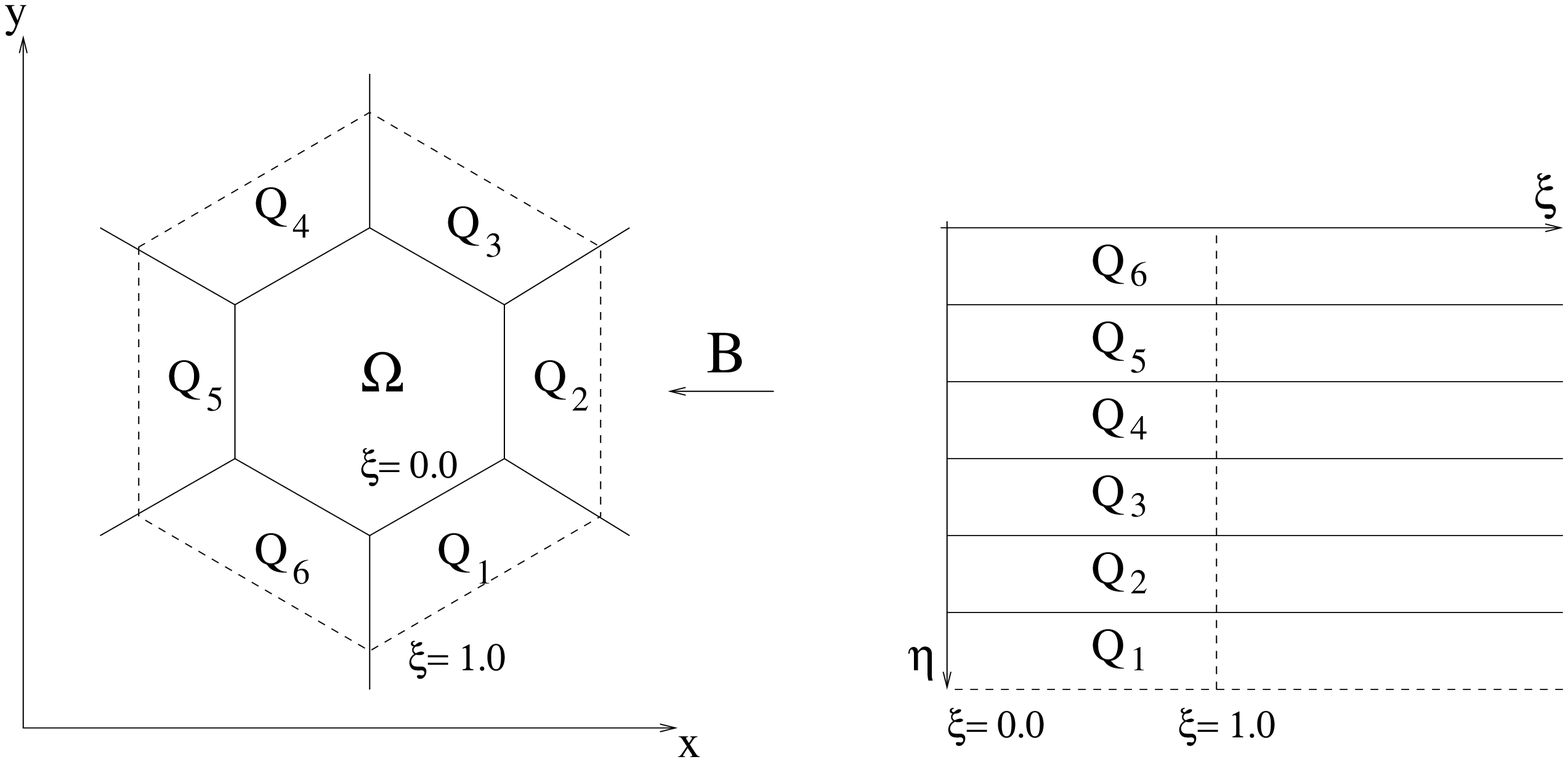}
    \end{tabular}
  \end{center}
  \caption{
\label{fig:pml_geometry2D}
Prismatoidal coordinate system. Each segment $Q_j$ is the image of 
a reference element under a bilinear mapping $B^{loc}_j$. These local mappings
are combined to a global mapping $B$ which is continuous in $\eta$.}
\end{figure} 
\label{weakwaveguideproblem}
Find $k_{z}$ such that there exists a field $\VField{e} \in \hclkz$ which 
is purely outgoing in the cross section and which satisfies 
\begin{subequations}
\label{wgequationdfweak}
\begin{eqnarray}
\int_{\rnum^{2}} \left( \mu^{-1} d_{1} \VField{e} \wedge d_{1} \overline{\VField{v}} - 
\omega^{2} (\epsilon \VField{e}) \wedge \overline{\VField{v}} \right ) & = & 0 \\
\int_{\rnum^{2}}  (\epsilon \VField{e}) \wedge d_{0} \overline{p} & = & 0
\end{eqnarray}
\end{subequations}
for any $\VField{v} \in \hclkz$, $p \in \hplkz$ with compact support in $x$ and $y$.
\end{problem}  
\section{LEAKY MODES AND OUTGOING BOUNDARY CONDITION}
\label{sect:trueandleaky}
We now address the definition {\em purely outgoing} in Problem~\ref{weakwaveguideproblem}. From a physical point of view, any propagating mode is admissible as long as there is no energy transport in the cross section from infinity. As mentioned in the introduction to this paper we want to define the transparent boundary condition with the help of the pole condition concept~\cite{SchmidtF:01b}, which we now detail for the one dimensional case. \\
 Let us assume that the permittivity and permeability are only dependent on $x$, $\epsilon = \epsilon(x)$, $\mu = \mu(x)$ and are constant in the right exterior domain $I_{+} = [0, +\infty)$. Then a TE mode satisfies the Helmholtz equation 
\begin{eqnarray*}
\label{HelmholtzEquation1D}
-\partial_{xx} E_y(x) + k_z^2E_y(x) -\omega^2 \mu \epsilon E_y(x) = 0,\; x \in I_{+} 
\end{eqnarray*}     
with general solution 
\begin{eqnarray*}
E_y = A e^{i \sqrt{\omega^2 \mu \epsilon -k_z^2}x} + B e^{-i \sqrt{\omega^2 \mu \epsilon -k_z^2}x}.
\end{eqnarray*}
If we define the square root so that $\Re{\sqrt{\omega^2 \mu \epsilon -k_z^2}}>0$ the first part is an outgoing wave and the second part is an incoming wave. Therefore, as ''physical'' boundary condition we must enforce $B=0$. 
\newpage
Let us regard the Laplace transform of $E_y$, 
\begin{eqnarray*}
\LT E_y = \int_0^\infty E_y(x)e^{-sx}\dd{x} = \frac{A}{s-i \sqrt{\omega^2 \mu \epsilon -k_z^2}}+
\frac{B}{s+i \sqrt{\omega^2 \mu \epsilon -k_z^2}}.
\end{eqnarray*}
We see that the incoming wave produces a pole at $s = -i \sqrt{\omega^2 \mu \epsilon -k_z^2}$. Hence $B=0$ is equivalent to the fact that the Laplace transform of the solution is holomorphic in the lower complex half plane. This is precisely the pole condition for the one dimensional case: 

{\em A solution to Helmholtz equation~\eqref{HelmholtzEquation1D} is purely outgoing if its Laplace transform is holomorphic in the lower complex half plane.}

To state the pole condition for the two dimensional case we map the exterior domain $\Omega_{\rm ext} \subset \rnum^{2}$ onto $\Omega_{\eta, \xi}$ as depicted in Figure~\ref{fig:pml_geometry2D}. Here we assume that the material properties are constant on each segment $Q_{j}$ but may vary from segment to segment. The $z$~-~coordinate remains unchanged under the transformation. The transformed Maxwell's equations are exactly of the form ~\eqref{thefieldmaxequationsdf} but with transformed tensors $\epsilon_{\eta, \xi}$ and $\mu_{\eta_, \xi}$. With the usual notation $e_{*}$ for the {\em pulled back} differential form the weak waveguide problem with transformed exterior domain now reads  
\begin{problem}[Weak Waveguide Problem with Transformed Exterior Domain]
\label{weakwaveguideproblemtransextdomain}
Find $k_{z}$ such that there exist fields $\VField{e}(x, y, z) \in \hckzd{\Omega_{\rm int}}$ and $\VField{e}_{*}(\eta, \xi, z) \in \hclkz$ such that:
\begin{enumerate}
\item
 $(\VField{e}_{*})^* = \VField{e}$ on the boundary $\partial \Omega$. (Matching~Condition) 
\item
$\widehat{\VField{e}}_{*}(\eta, s)  = \LT \VField{e}_{*} (\eta, \xi)$ defines a holomorphic function on the lower complex half plane ($\Im{s} \leq 0$). (Pole~Condition)
\item The field composed of $\VField{e}$ and $\VField{e_{*}}^{*}$ satisfies Maxwell's equations:
\begin{subequations}
\begin{eqnarray*}
\int_{\Omega_{\rm int}} \left( \mu^{-1} d_{1} \VField{e} \wedge d_{1} \overline{\VField{v}} - 
\omega^{2} (\epsilon \VField{e}) \wedge \overline{\VField{v}} \right ) +
\int_{\Omega_{\eta, \xi}}  \left( \mu^{-1}_{\eta, \xi} d_{1} \VField{e}_{*} \wedge d_{1} \overline{\VField{v}}_{*} - 
\omega^{2} (\epsilon_{\eta, \xi} \VField{e}_{*}) \wedge \overline{\VField{v}}_{*} \right ) 
& = & 0 \\
\int_{\Omega_{\rm int}}  (\epsilon \VField{e}) \wedge d_{0} \overline{p} +
\int_{\Omega_{\eta, \xi}}  (\epsilon_{\eta, \xi} \VField{e}_{*} \wedge d_{0} \overline{p}_{*}) 
& = & 0
\end{eqnarray*}
\end{subequations}
for any $\VField{v} \in \hckzd{\Omega_{\rm int}}$, $p \in \hpkzd{\Omega_{\rm int}}$, and compactly supported $\VField{v}_{*} \in \hclkz$, $p_{*} \in \hplkz$ such that $(\VField{v}_{*})^{*} = \VField{v}$, $(p_{*})^{*} = p$ on the boundary $\xi = 0$.
\end{enumerate}
\end{problem}  
\section{TRANSPARENT BOUNDARY CONDITIONS}
\label{sect:transparentboundary} 
\begin{figure}
  \begin{center}
    \begin{tabular}{c}
      \includegraphics[width=12cm]{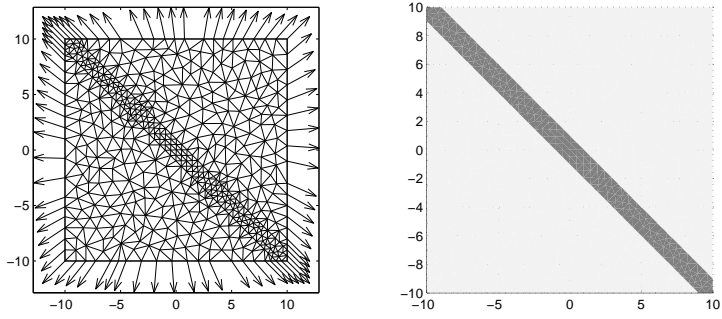}
    \end{tabular}
  \end{center}
  \caption{Discretization of the interior domain and rays in the exterior domain
    (left picture). Geometry with representation of the refractive index 
    distribution (right picture). Infinite waveguide: $k_{2}=1.32$, background: 
    $k_{1}=0.29$.}
  \label{Fig:TMWaveguide}
\end{figure}
Problem~\ref{weakwaveguideproblemtransextdomain} is still posed on an unbounded domain and therefore numerically not feasible. As mentioned in the introduction to this paper the transformed exterior field $\VField{e}_{*}$ is typically not decreasing in the exterior domain. This rules out a simple truncation of the computational domain. When constructing transparent boundary conditions the aim is to compute the true solution in the interior domain with a numerical effort proportional to the number of unknowns in the interior domain. As shown by Schmidt et al.~\cite{SchmidtF:01b,Hohage:03b} the Laplace transform $\widehat{\VField{e}}_{*}$ behaves very kindly. As numerically approved, a discretization of $\widehat{\VField{e}}_{*}$ along the real axis with global functions gives a transparent boundary condition so that the computed interior solution  converges exponentially fast to the true solution (up to the interior discretization error) with the number of discretization ``points'' used for $\widehat{\VField{e}}_{*}$. \\

In this paper we focus on the Perfectly Matched Layer method introduced by Berenger~\cite{Berenger:94a,Berenger:95a,Collino:98a}. To motivate the method we go back to the one dimensional Helmholtz equation~\eqref{HelmholtzEquation1D}. The general solution in the exterior domain $I_{+}$ is holomorphic in $x$. We see that along the straight line $(1+i\sigma)$ the outgoing part becomes exponentially decreasing as far as $\sigma$ is chosen such that $\sigma|Re\sqrt{\omega^2 \mu \epsilon -k_z^2}| > |\Im\sqrt{\omega^2 \mu \epsilon -k_z^2}|$ while the incoming field explodes, 
\begin{eqnarray*}
E_y = A \underbrace{e^{i \sqrt{\omega^2 \mu \epsilon -k_z^2}(1+i\sigma)x}}_{\mbox{outgoing} \, \thicksim \, \mbox{evanescent}} + B \underbrace{e^{-i \sqrt{\omega^2 \mu \epsilon -k_z^2}(1+i\sigma)x}}_{\mbox{incoming} \, \thicksim \, \mbox{exploding}}.
\end{eqnarray*}
Imposing now a zero Dirichlet boundary condition at $x=\rho$ and assuming that the field intensity is equal to one for $x=0$ yields 
\[
|B| = \left | \frac{e^{i \sqrt{\omega^2 \mu \epsilon -k_z^2}(1+i\sigma)x}}{e^{i \sqrt{\omega^2 \mu \epsilon -k_z^2}(1+i\sigma)x} +e^{-i \sqrt{\omega^2 \mu \epsilon -k_z^2}(1+i\sigma)x}} \right | \sim e^{-\Re{\sqrt{\omega^2 \mu \epsilon -k_z^2}} \sigma \rho}. 
\]
Therefore, the true boundary condition $B=0$ is enforced exponentially fast with the layer thickness $\rho$.   

In order to switch to the higher dimensional case we assume that $\VField{e}_{*}(\eta, \xi)$ possesses a holomorphic extension in $\xi.$ For a homogeneous exterior domain and some special inhomogeneous exterior domains this is proved in Hohage et al.~\cite{Hohage:03c}. It is an aim for the future work of the authors to prove that in general a field $\VField{e}_{*}$ satisfying the pole condition also has an holomorphic extension in $\xi$. For $\gamma=1+i\sigma$ let us denote  $\VField{e}_{*, {\rm B}}(\eta, \xi) = \VField{e}_{*}(\eta, \gamma \xi)$, $\epsilon_{\eta, \xi, {\rm B}}(\eta, \xi) = \epsilon_{\eta,  \xi}(\eta, \gamma \xi)$, and $\mu_{\eta, \xi, {\rm B}}(\eta, \xi) = \mu_{\eta,  \xi}(\eta, \gamma \xi)$. The holomorphic extension $\VField{e}_{*, {\rm B}}(\eta, \xi)$ is called Berenger function. One expects that the field $\VField{e}_{*, {\rm B}}(\eta, \xi)$ decays exponentially fast for $\xi \rightarrow \infty$. Again this admits to truncate the computational domain to $\Omega_{\rm PML} = [\eta_{\rm min}, \eta_{\rm max}] \times [0, \rho)$ and to impose a zero Dirichlet boundary condition at $\xi = \rho$. We are lead to the following PML problem where $d_{k, \rm B}$ denotes the exterior derivative with $\partial_{\xi}$ replaced by $1/(1+i\sigma)\partial_{\xi}$

\begin{problem}[Weak Waveguide Problem with PML]
\label{weakwaveguideproblempml}
Find $k_{z}$ such that there exist fields $\VField{e}(x, y, z) \in \hckzd{\Omega_{\rm int}}$ and $\VField{e}_{*, {\rm PML}}(\eta, \xi, z) \in \hckzd{\Omega{\rm PML}}$ such that $(\VField{e}_{*})^* = \VField{e}$ on the boundary $\partial \Omega$ (Matching~Condition) and 
\begin{subequations}
\begin{eqnarray*}
\int_{\Omega_{\rm int}} \left( \mu^{-1} d_{1} \VField{e} \wedge d_{1} \overline{\VField{v}} - 
\omega^{2} (\epsilon \VField{e}) \wedge \overline{\VField{v}} \right ) +
\gamma \int_{\Omega_{\eta, \xi}, \rho}  \left( \mu^{-1}_{\eta, \xi, {\rm B}} d_{1, {\rm B}} \VField{e}_{*, {\rm PML}} \wedge d_{1} \overline{\VField{v}}_{*} - 
\omega^{2} (\epsilon_{\eta, \xi, {\rm B}} \VField{e}_{*, {\rm PML}}) \wedge \overline{\VField{v}}_{*)} \right ) 
& = & 0 \\
\int_{\Omega_{\rm int}}  (\epsilon \VField{e}) \wedge d_{0} \overline{p} +
\gamma \int_{\Omega_{\eta, \xi, \rho}}  (\epsilon_{\eta, \xi, {\rm B}} \VField{e}_{*, {\rm PML}} \wedge d_{0, {\rm B}} \overline{p}_{*}) 
& = & 0 \\
\VField{e}_{*, {\rm PML}}|_{\xi = \rho} & = & 0
\end{eqnarray*}
\end{subequations}
for any $\VField{v} \in \hckzd{\Omega_{\rm int}}$, $p \in \hpkzd{\Omega_{\rm int}}$ and  $\VField{v}_{*} \in \hckzd{\Omega_{\rm PML}}$, 
$p_{*} \in \hpkzd{\Omega_{\rm PML}}$  such that $(\VField{v}_{*})^{*} = \VField{v}$, $(p_{*})^{*} = p$ on the boundary $\partial \Omega_{\rm int}$.
\end{problem}  

\begin{remark}
The complex continuation along the straight line $\gamma \xi$ yields a jump in the Neumann boundary condition at $\xi=0$,
\[
\int_{\xi=0}  \mu^{-1}_{\eta, \xi, {\rm B}} d_{1, {\rm B}} \VField{e}_{*, {\rm B}} \wedge \overline{\VField{v}}_{*} = \gamma 
\int_{\xi=0}  \mu^{-1}_{\eta, \xi} d_{1} \VField{e}_{*} \wedge \overline{\VField{v}}_{*}.
\]
The factor $\gamma$ left of the integral symbols $\int_{\eta, \xi, \rho}$ in Problem~\ref{weakwaveguideproblempml} is introduced to incorporate this jump in the variational problem as the natural boundary condition on $\partial \Omega_{\rm int}$. This avoids the definition of further unknowns on the boundary (Lagrange parameters).  
\end{remark}

\begin{figure}
  \begin{center}
    \begin{tabular}{c}
      \includegraphics[width=6cm]{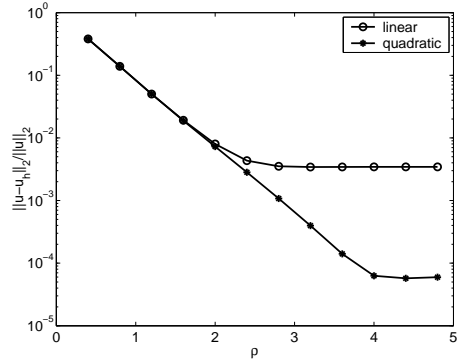}
    \end{tabular}
  \end{center}
\caption{Relative error $||u-u_h||_2/||u||_2$ versus thickness of the 
PML-layer 
for  linear and quadratic finite elements in the first experiment.  }
\label{Fig:WaveguideTMConvergence}
\end{figure}
The PML method is proved to converge exponentially fast to the true solution with an increasing layer thickness $\rho$ for a homogeneous exterior domain~\cite{Lassas:98a,Lassas:01a} and for some special inhomogeneous exterior domains~\cite{Hohage:03c}. To demonstrate the accuracy and the exponential convergence of the method even for rather complex exterior domains we want to compute the propagation of a TM polarized fundamental mode,
\[
-\Delta E_{z} - k^{2}(x, y)E_{z} = 0
\]
along a waveguide as depicted in Figure~\ref{Fig:TMWaveguide}, see also Zschiedrich et al.~\cite{Zschiedrich:2003a} The fundamental mode is used as an incoming field and is only specified along the left and upper side of the computational domain. Thus this example is a non trivial {\rm scattering problem} - we must recover the propagating mode in the interior domain. The exterior domain is non-homogeneous due to the infinite waveguide. Figure~\ref{Fig:WaveguideTMConvergence} shows the relative $L_{2}$~-~error in the computational domain. We observe exponential convergence for growing thickness $\rho$ of the PML layer until the discretization error of the interior problem dominates the overall error.    
\section{FINITE ELEMENT DISCRETIZATION}
\label{sect:finiteelements}
To discretize Problem~\ref{weakwaveguideproblempml} we split the interior field 
\[
\VField{e} = e_{x}(x,y)e^{ik_{z}z}dx+e_{y}(x, y)e^{ik_{z}z}dy+e_{z}(x, y)e^{ik_{z}z}dz
\]
into a transversal part $\VField{e}_{\perp} = e_{x}(x,y)dx+e_{y}(x, y)dy$ and a longitudinal part $\VField{e}_{z} = e_{z}(x, y)dz$. As usual we discretize $\VField{e}_{\perp}$ with Nedelec's edge elements and $e_{z}$ with standard scalar elements. This gives a discrete counterpart to the de Rham complex and hence leads to a discrete divergence condition~\cite{Beck:96a}. In this way, spurious modes which may rise from the kernel of the $\nabla \times$~-~operator when using an improper discretization scheme are ruled out. The variational problem for the interior problem reads in classical notation  
\begin{eqnarray*}
\label{weakpropeigenproblem}
\int_{\rnum^2} \mu^{-1}
\left[ \begin{array}{c} \nabla E_z-ik_z\VField{E}_{\perp} \\ \nabla_{\perp} \times \VField{E}_\perp \end{array} \right] \cdot
\left[ \begin{array}{c} \nabla v_z^*-ik_z\VField{v}_{\perp}^* \\ \nabla_{\perp} \times \VField{v}_\perp^* \end{array} \right]-\omega^2 \epsilon
\left[ \begin{array}{c} \VField{E}_{\perp} \\ E_z \end{array} \right] \cdot
\left[ \begin{array}{c} \VField{v}_{\perp}^* \\ v_z^* \end{array} \right] \dd{x} \dd{y} & = & 0, \\
\int_{\rnum^2} \epsilon
\left[ \begin{array}{c} \VField{E}_{\perp} \\ E_z \end{array} \right] \cdot
\left[ \begin{array}{c} \nabla v_{z}^* \\ i k_z v_z^* \end{array} \right] \dd{x} \dd{y}& = & 0,
\end{eqnarray*}
for all $\VField{v}_{\perp} \in  H_{\Omega_{\rm int}}({\rm curl})$ and $v_z \in H^{1}_{\Omega_{\rm int}}$ 
with operators
\begin{eqnarray*}
\nabla_{\perp}  \,:\, H^{1}_{\Omega_{\rm int}} & \rightarrow & H_{\Omega_{\rm int}}({\rm curl}) \\
\phi & \mapsto & (\partial_x \phi, \partial_y \phi), 
\end{eqnarray*}
\begin{eqnarray*}
\nabla_{\perp} \times  \,:\, H_{\Omega_{\rm int}}({\rm curl}) & \rightarrow &  H_{\Omega_{\rm int}}({\rm div})\\
\VField{E}_{\perp} & \mapsto & (\partial_x E_y - \partial_y E_x), 
\end{eqnarray*}
and
\begin{eqnarray*}
\nabla_{\perp} \cdot  \,:\, H_{\Omega_{\rm int}}({\rm div}) & \rightarrow & L^{2} \\
\VField{D}_{\perp} & \mapsto & (\partial_x D_x + \partial_y D_y), 
\end{eqnarray*}
Again one sees that any solution to the first equation also solves the second one (divergence condition). Simply set $\VField{v}_{\perp} = 1/(ikz) \nabla v_z$ for $k_z \neq 0$ and recall that $\nabla_{\perp} \times \nabla_{\perp} = 0$. For $k_z=0$ set $v_z=0$ and $v_\perp = \nabla p$ for any $p \in H^{1}_{\Omega_{\rm int}}$. 
Within the PML layer we use corresponding finite elements on quadrilaterals, which are defined on a reference quadrilateral via a tensor product ansatz~\cite{Zschiedrich:2003a}. On the whole transformed exterior domain $\Omega_{\rm PML}$ we use a fixed discretization in $\xi$~-~direction. For the interior discretization we have implemented an adaptive grid refinement steered by a residual based error estimator as in Heuveline and Rannacher~\cite{Heuveline:2001a}.
\section{EXAMPLES}
We now demonstrate the ability of our code to cope with challenging problems in the optical waveguide design. In the examples, the quadratric eigenvalue problem is solved with the ARPACK package by Sorensen et al.~\cite{Lehoucq:97a} after a reduction to a linear eigenvalue problem. The Arnoldi method is used in the shift-invert mode and we rely on Intel's Math Kernel Library for sparse LU decomposition (PARDISO~\cite{Hagemann:2004a}). 
\label{sect:examples}
\subsection{Plasmon Polariton Mode}
\begin{figure}
  \begin{center}
    \begin{tabular}{c}
\psfrag{a}{$a$}
\psfrag{d1}{$d_{1}$}
\psfrag{d2}{$d_{2}$}
\psfrag{w}{$w$}
\includegraphics[width=6cm]{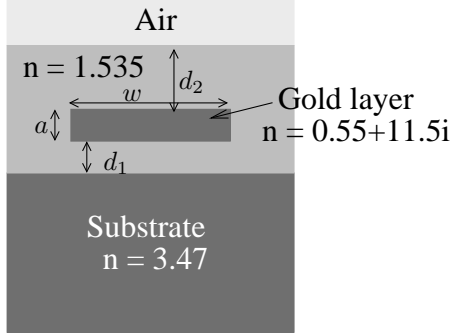}
\end{tabular}
\end{center}
\caption{
  \label{fig:plasmonscetch}
  Plasmon-Polariton-Waveguide. In the computations we have used $a=10nm$, $w=20 \mu m$, $d_{1}=4\mu m$ and $d_{2}=8.01\mu m$. }
\end{figure} 
\begin{figure}
  \begin{center}
    \begin{tabular}{cc}
\includegraphics[width=7cm]{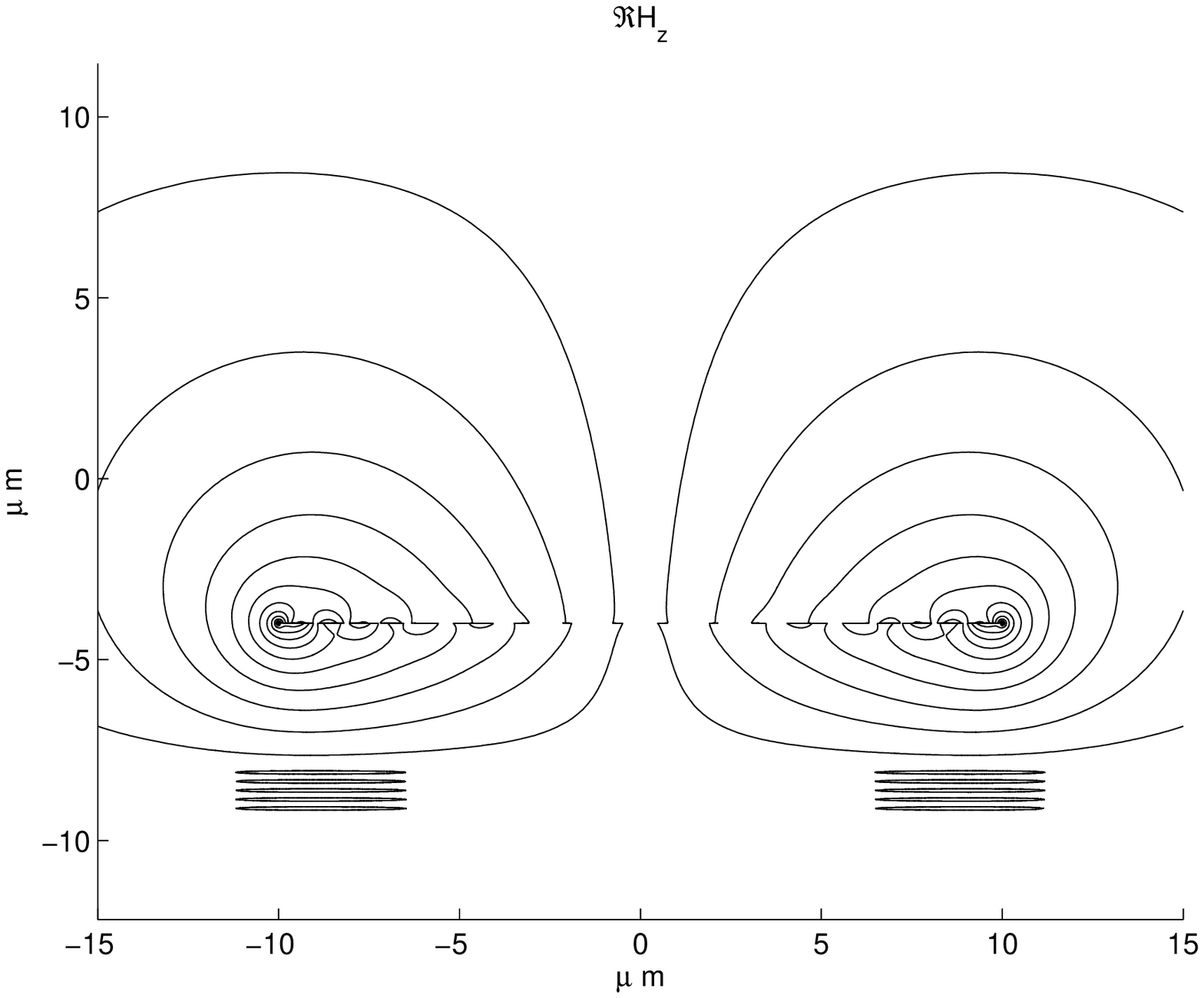} &
\includegraphics[width=7cm]{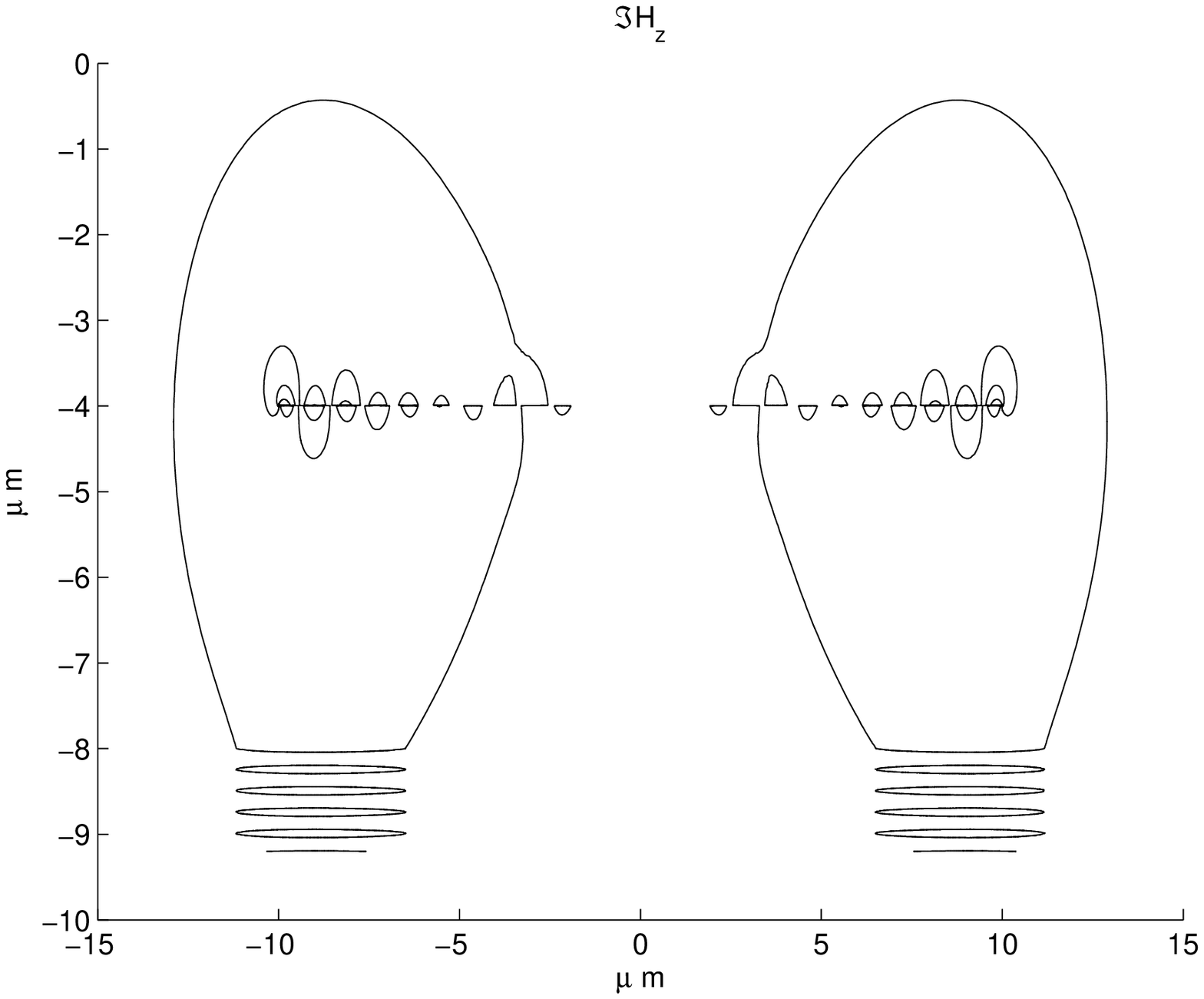}
\end{tabular}
\end{center}
\caption{
  \label{fig:plasmonz}
  Plasmon-Polariton-Waveguide. Real and imaginary parts of the $H_{z}$~-~component for the fundamental mode.}
\end{figure} 
\begin{figure}
  \begin{center}
    \begin{tabular}{cc}
\includegraphics[width=7cm]{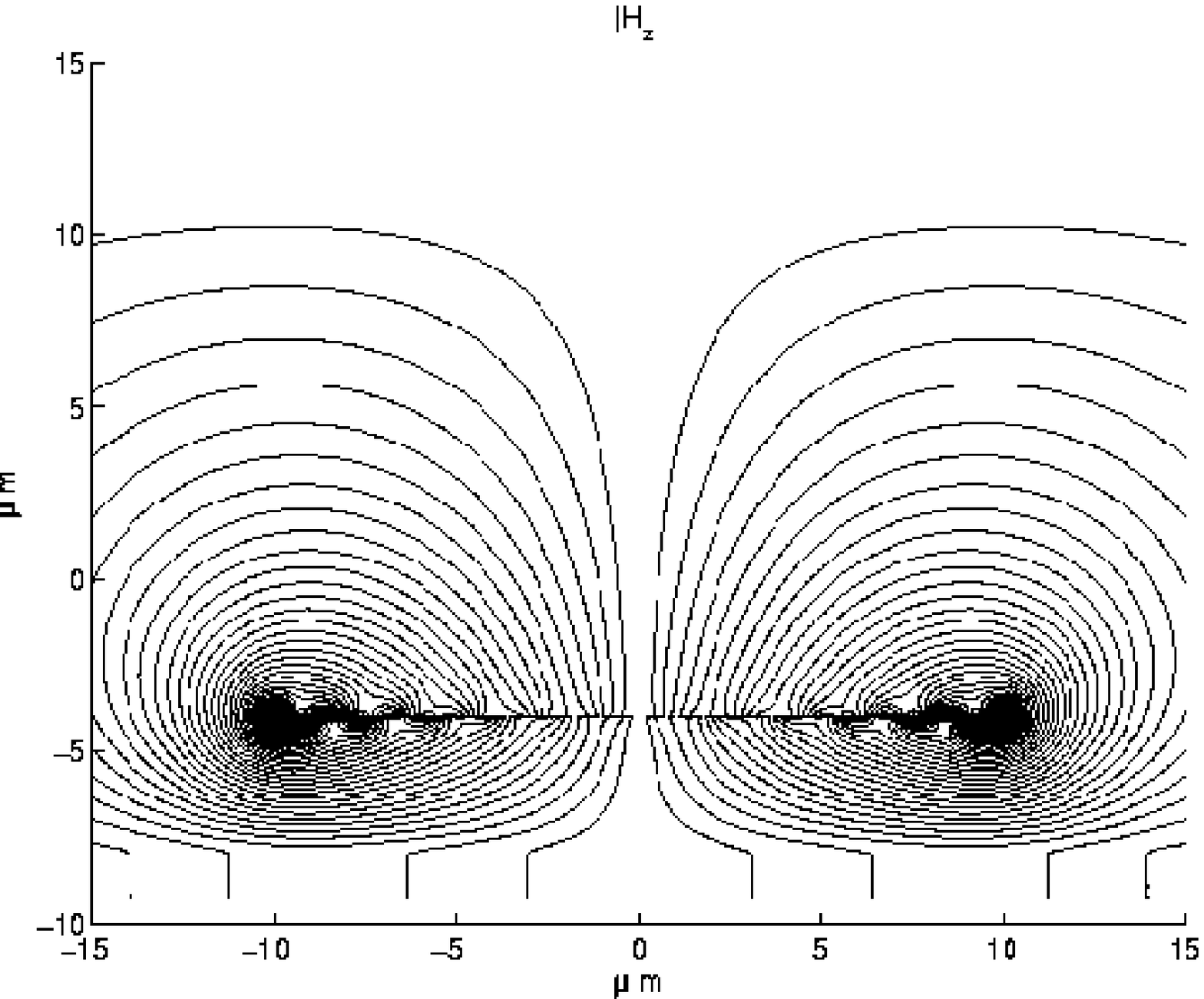} &
\includegraphics[width=7cm]{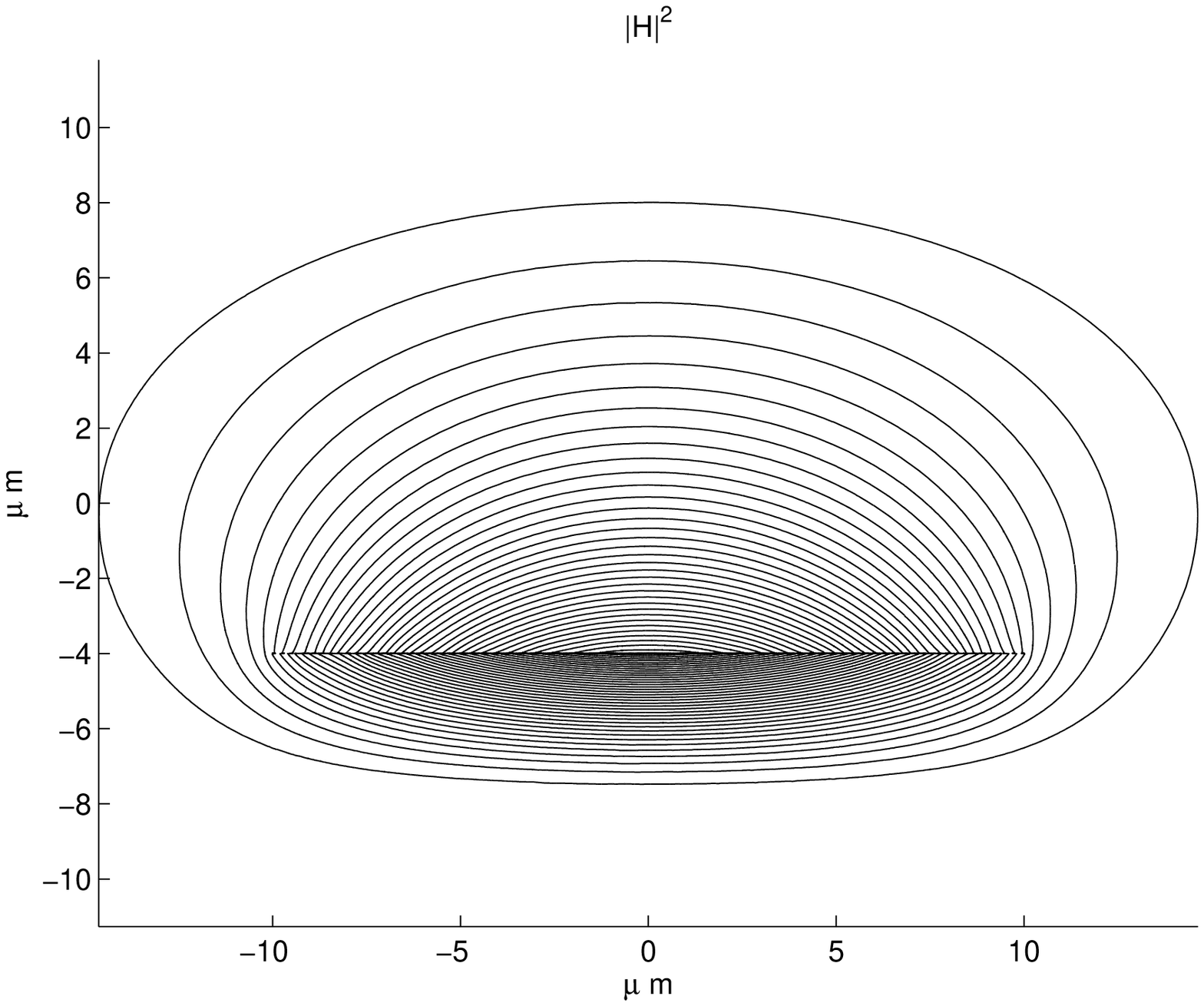}
\end{tabular}
\end{center}
\caption{
  \label{fig:plasmont}
  Plasmon-Polariton-Waveguide. Left: Absolute value of $H_{z}$~-~component for the fundamental mode. There appear no spurious reflections at the lower boundary. Right: Magnetic field intensity. The mode is localized near the metal stripe.}
\end{figure} 
As shown by Berini et al.~\cite{Berini:99a} and Bozhevolnyi~\cite{Bozhevolnyi:?a} a very thin metal stripe may serve as a waveguide. In this case the propagating mode is localized near the metal stripe. The present geometry is sketched in Figure~\ref{fig:plasmonscetch}. Since the substrate has a relatively high refractive index the modes are typically leaky. Further there are singularities near the metal's corner. This calls for an adaptive grid refinement. The coarse grid consists of 13684 triangles and is adaptively refined three times during the program execution. Within the PML layer we have used the discretization $\xi = $~[0.0 : 0.1 : 2.0].\^{ }3 (in Matlab notation). As the initial guess for $k_{z}$ we have used the result from the one dimensional problem which is given by a cut along the symmetry axis of the waveguide.  In Table~\ref{tab:plasmonruns} the computed effective refractive index for the fundamental mode and the computation effort are given. We observe convergence up to eight digits after three grid refinement steps. In Figures~\ref{fig:plasmonz} and~\ref{fig:plasmont} one sees isoline-plots for the magnetic field strength which show that there are no spurious reflections from the boundary of the computational domain.  
\begin{table}
\begin{center}
\begin{tabular}{rrrrr}
Step & $n_{PML, \rm eff}$ & $N^o$ DOF & total time [min] & Memory [GByte] \\
\hline
 0 &  1.5350262e+00+0.0000981e+00i  & 159729  &  01:57 & $\sim 0.9$ \\
 1 &  1.5350261e+00+0.0000985e+00i  & 281151  &  03:35 & $\sim 2.6$ \\ 
 2 &  1.5350263e+00+0.0000984e+00i  & 527656  &  07:52 & $\sim 4.8$ \\ 
 3 &  1.5350263e+00+0.0000984e+00i  & 881016  &  12:16 & $\sim 9.1$
\end{tabular}
\end{center}
\caption{Fundamental mode of the Plasmon Polariton waveguide for a vacuum wavelength of $\lambda_{0}=1.55 \mu m$. 
The computations were performed on an AMD Opteron Linux-PC. 
\label{tab:plasmonruns}}
\end{table}
\subsection{Arrow Waveguide}
\begin{figure}
  \begin{center}
    \begin{tabular}{cc}
\includegraphics[width=15cm]{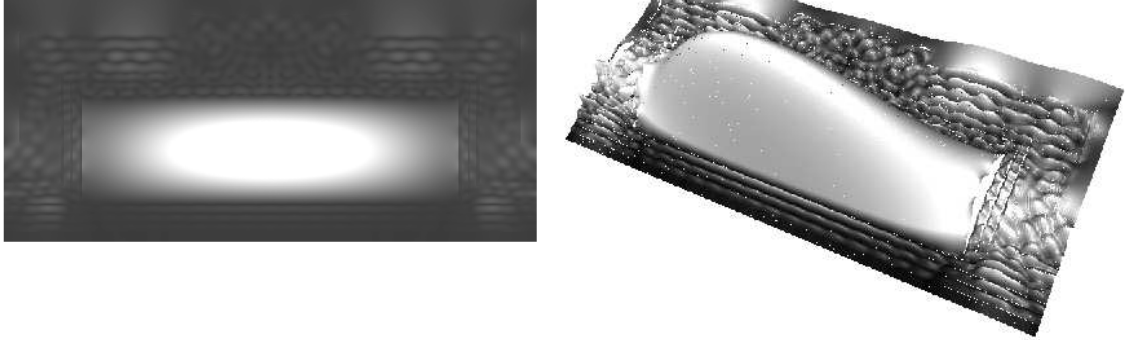} &
\end{tabular}
\end{center}
\caption{Fundamental leaky mode of the studied Arrow waveguide: Magnitude of the electric field, $|E(x, y)|$ in the cross section. The gray colormap in the right part of the figure is lighted up so that the field in the Arrow layers is better visible. Recall that the normal component of the electric field jumps across material boundaries.
  \label{fig:arrowt}}
\end{figure} 
\begin{table}
\begin{center}
\begin{tabular}{rrr|rr}
Step & $n_{PML, \rm eff}$ & $N^o$ DOF & $n_{PML, \rm eff}$ & $N^o$ DOF\\
\hline
 0 &  9.9325021e-01+0.0012272e-01i  &  51111 & 9.9325021e-01+0.0012272e-01i  &  51111 \\
 1 &  9.9322697e-01+0.0017419e-01i  &  92260 & 9.9322724e-01+0.0017697e-01i & 135625  \\ 
 2 &  9.9320708e-01+0.0016724e-01i  & 154747 & 9.9320699e-01+0.0017118e-01i & 404865 \\
 3 &  9.9320222e-01+0.0016547e-01i  & 265375 & 9.9320499e-01+0.0016816e-01i & 1344193 \\
 4 &  9.9320574e-01+0.0016710e-01i  & 478785 & {} & {} \\
 5 &  9.9320580e-01+0.0016820e-01i  & 1449444 & {} & {}
\end{tabular}
\end{center}
\caption{Fundamental leaky mode of the Arrow waveguide for a vacuum wavelength of $\lambda_{0}=785nm$. The left part corresponds to an adaptive grid refinement, the right part to a uniform grid refinement.
The computations were performed on an AMD Opteron Linux-PC. Computation time and memory requirements are similar to the previous example for a equal number of unknowns. Observe that with an adaptive refinement strategy the memory requirements remain below the 32-bit PC limit up to the third adaptive refinement step.
\label{tab:arrowwaveguideruns}}
\end{table}
\begin{figure}
  \begin{center}
    \begin{tabular}{cc}
      \includegraphics[width=5cm]{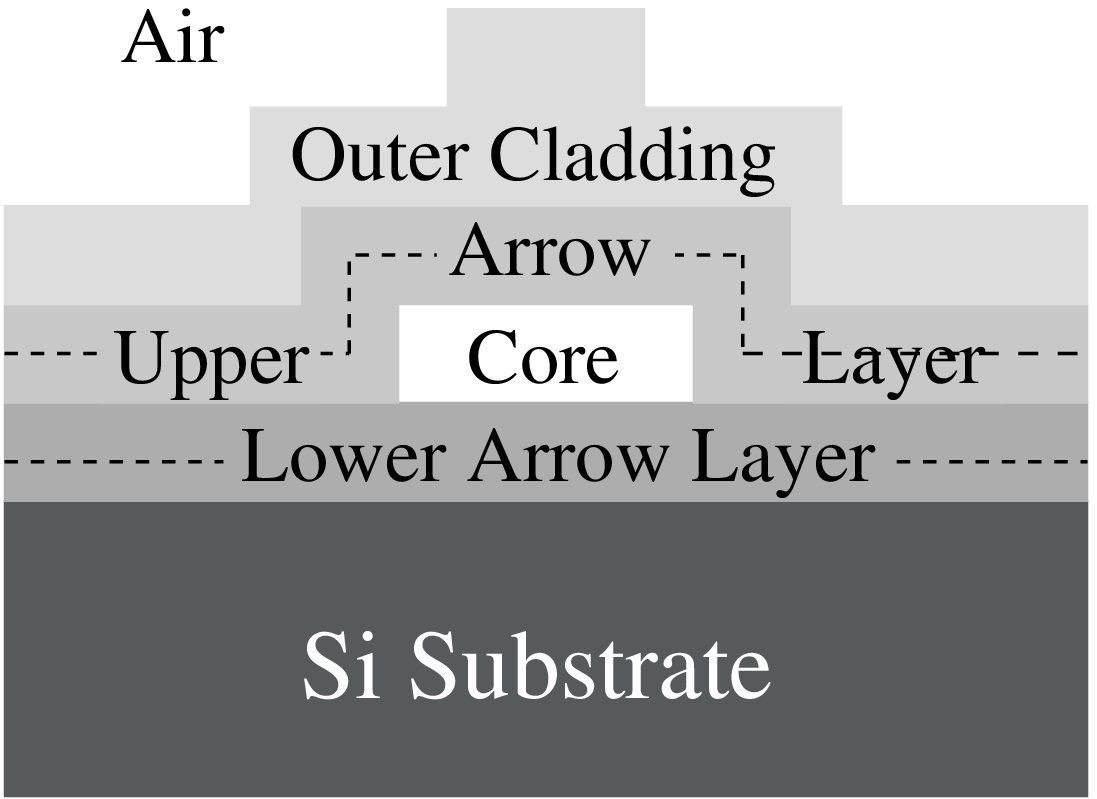} \vspace{0.5cm}&\vspace{0.5cm}
\psfrag{n1}{$n_{1}$}
\psfrag{n2}{$n_{2}$}
\psfrag{w1}{$w_{1}$}
\psfrag{w2}{$w_{2}$}
      \includegraphics[width=5cm]{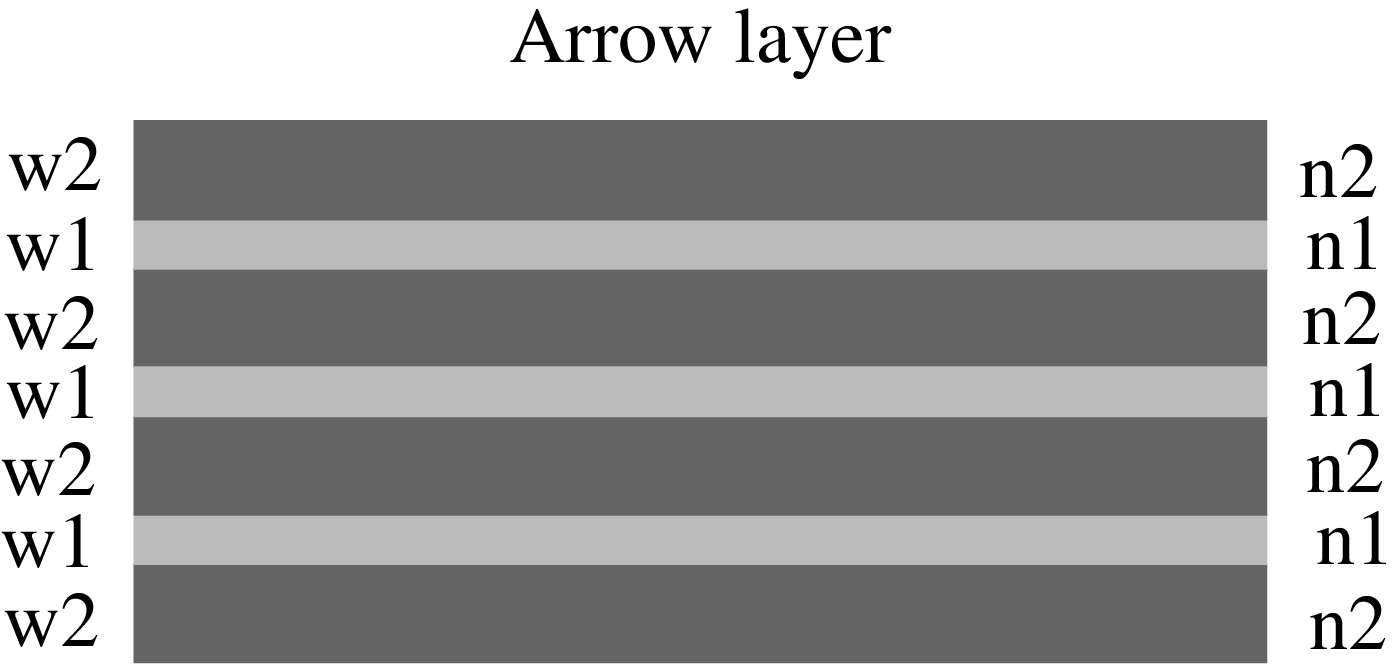} 
    \end{tabular}
  \end{center}
  \caption{
\label{fig:scetcharrow}
Hollow core ARROW waveguide. The core width is equal to $12\mu m$ and the core height is equal to $3.5 \mu m$. The Arrow layers are composed of silicon nitride ($n_{1}=2.1$, $w_{1} = 109nm$) and silicon oxide ($n_{2}=1.46$, $w_{2} = 184nm$) the substrate has a refractive index of $n = 3.4975$.}
\end{figure} 
The present waveguide structure consists of a hollow, rectangular core investigated in Yin et al.~\cite{Yin:2004a}. The field is confined by antiresonant, reflecting optical layers (ARROW). The geometry is sketched in Figure~\ref{fig:scetcharrow}. Again as an initial guess we have used the results from the corresponding one dimensional problem on the cut along the symmetry axis of the waveguide. Interestingly without transparent boundary conditions we were not able to find the two dimensional fundamental mode with primarily TE-polarization. Figure~\ref{fig:arrowt} shows the magnitude of the fundamental mode. In Table~\ref{tab:arrowwaveguideruns} the computed effective refractive index for the fundamental mode is given. The adaptive grid refinement allows to compute the propagation mode with a reasonable accuracy even on a 32-bit PC. 
\section*{ACKNOWLEDGMENTS}       
We thank P. Deuflhard and R. M{\"a}rz for fruitful discussions, and we acknowledge support by the initiative DFG Research Center {\em Matheon} of the Deutsche Forschungsgemeinschaft, DFG, and by the German Federal Ministry of Education and Research, BMBF, under contract no. 13N8252 ({\em HiPhoCs}). 
\bibliography{/home/numerik/bzfzschi/latex/Allgemein/Literatur}   
\bibliographystyle{spiebib}  
\end{document}